\documentclass[11pt]{article}

\usepackage[letterpaper,margin=1in,headheight=16pt]{geometry}
\usepackage[T1]{fontenc}
\usepackage{lmodern}
\usepackage{graphicx}
\usepackage{booktabs}
\usepackage{enumitem}
\usepackage{titlesec}
\usepackage{fancyhdr}
\usepackage{amsmath}
\usepackage{amssymb}
\usepackage[numbers,sort&compress]{natbib}

\newcommand{\ergs}{\ifmmode \textrm{erg\ s}^{-1} \else erg s$^{-1}$\fi}
\newcommand{\Msun}{\ifmmode \textrm{M}_{\odot} \else $M_{\odot}$\fi}
\newcommand{\Lsun}{\ifmmode \textrm{L}_{\odot} \else $L_{\odot}$\fi}

\usepackage{xcolor}
\definecolor{genesisgold}{HTML}{B5771A}
\definecolor{genesisink}{HTML}{10131F}
\definecolor{genesisgray}{HTML}{6B7386}

\usepackage[colorlinks=true,linkcolor=genesisgold,citecolor=genesisgold,urlcolor=genesisgold]{hyperref}

\titleformat{\section}{\normalfont\large\bfseries\color{genesisink}}{}{0pt}{}
\titlespacing*{\section}{0pt}{1.3em}{0.45em}

\fancypagestyle{genesisfirst}{%
  \fancyhf{}%
  \fancyhead[L]{}%
  \fancyhead[R]{\small\textcolor{genesisgray}{JWST MCT: black hole seeds and LRDs}}%
  \fancyfoot[C]{\small\thepage}%
}

\newcommand{\Mblack}{M_\bullet}
\newcommand{\Mstar}{M_\star}

\begin{document}

\newcommand{\AuthorList}{\begingroup\sloppy\emergencystretch=3em
\mbox{F.~Pacucci\textsuperscript{1,\dag}},
\mbox{R.~Schneider\textsuperscript{2,3,\dag}},
\mbox{H.~Katz\textsuperscript{4}},
\mbox{B.~Robertson\textsuperscript{5}},
\mbox{R.~Somerville\textsuperscript{6}},
\mbox{Abdurro'uf\textsuperscript{7}},
\mbox{K.~Aditya\textsuperscript{8}},
\mbox{J.~Afonso\textsuperscript{9,10}},
\mbox{J.~Arita\textsuperscript{11}},
\mbox{G.~Barro\textsuperscript{12}},
\mbox{A.~Basu\textsuperscript{13}},
\mbox{R.~Begley\textsuperscript{14}},
\mbox{S.~Bose\textsuperscript{15}},
\mbox{M.~Brada\v{c}\textsuperscript{16,17}},
\mbox{G.~Bryan\textsuperscript{18}},
\mbox{C.~Cain\textsuperscript{19,20,21}},
\mbox{R.~Caleno\textsuperscript{2,3,22}},
\mbox{A.~Cameron\textsuperscript{23,24}},
\mbox{S.~Carniani\textsuperscript{25}},
\mbox{H.~Choi\textsuperscript{26}},
\mbox{N.~Choustikov\textsuperscript{19,20}},
\mbox{E.~Curtis-Lake\textsuperscript{27}},
\mbox{A.~D'Angelo\textsuperscript{3,28,2,22}},
\mbox{R.~Dave\textsuperscript{29}},
\mbox{R.~Davies\textsuperscript{30}},
\mbox{K.~Davis\textsuperscript{31,32}},
\mbox{C.~Donnan\textsuperscript{33}},
\mbox{Q.~Duan\textsuperscript{19}},
\mbox{Y.~Dubois\textsuperscript{34}},
\mbox{A.~Faisst\textsuperscript{35}},
\mbox{M.~Farcy\textsuperscript{36}},
\mbox{S.~Finkelstein\textsuperscript{37,38}},
\mbox{S.~Gillman\textsuperscript{39}},
\mbox{S.~Glover\textsuperscript{67}},
\mbox{L.~Graziani\textsuperscript{2}},
\mbox{A.~Gupta\textsuperscript{37,38}},
\mbox{T.~Gutcke\textsuperscript{40}},
\mbox{M.~Haehnelt\textsuperscript{20,19}},
\mbox{N.~Hathi\textsuperscript{41}},
\mbox{E.~Hatziminaoglou\textsuperscript{42,43}},
\mbox{K.~Heintz\textsuperscript{39}},
\mbox{M.~Hirschmann\textsuperscript{36}},
\mbox{L.~C.~Ho\textsuperscript{70}},
\mbox{L.~Ivey\textsuperscript{19}},
\mbox{K.~Iyer\textsuperscript{40}},
\mbox{J.~Jeon\textsuperscript{37,38}},
\mbox{B.~Jones\textsuperscript{44}},
\mbox{I.~Juod\v{z}balis\textsuperscript{45}},
\mbox{A.~Koekemoer\textsuperscript{46}},
\mbox{C.~Kragh Jespersen\textsuperscript{47}},
\mbox{D.~Langeroodi\textsuperscript{19}},
\mbox{A.~Lapi\textsuperscript{48}},
\mbox{M.~Latif\textsuperscript{49}},
\mbox{Y.~Liu\textsuperscript{15}},
\mbox{C.~Lovell\textsuperscript{19}},
\mbox{F.~Mannucci\textsuperscript{68}},
\mbox{K.~Nagamine\textsuperscript{50}},
\mbox{P.~Natarajan\textsuperscript{51,52,53}},
\mbox{Z.~Pan\textsuperscript{54}},
\mbox{R.~Pascalau\textsuperscript{19}},
\mbox{P.~G.~P\'erez-Gonz\'alez\textsuperscript{55}},
\mbox{E.~Pizzati\textsuperscript{1}},
\mbox{E.~Regos\textsuperscript{56}},
\mbox{P.~Rinaldi\textsuperscript{37,38}},
\mbox{D.~Roberts\textsuperscript{57}},
\mbox{W.~Roper\textsuperscript{69}},
\mbox{D.~Schleicher\textsuperscript{2}},
\mbox{F.~Shankar\textsuperscript{57}},
\mbox{X.~Shen\textsuperscript{1}},
\mbox{T.~Stanton\textsuperscript{29}},
\mbox{G.~Sun\textsuperscript{58}},
\mbox{Y.~Sun\textsuperscript{59}},
\mbox{A.~Trinca\textsuperscript{3}},
\mbox{J.~Trump\textsuperscript{31}},
\mbox{J.~Turner\textsuperscript{60}},
\mbox{H.~Übler\textsuperscript{61}},
\mbox{R.~Valiante\textsuperscript{3}},
\mbox{F.~van Donkelaar\textsuperscript{20,19}},
\mbox{G.~Venturi\textsuperscript{68}},
\mbox{A.~Vijayan\textsuperscript{60}},
\mbox{F.~Vito\textsuperscript{62}},
\mbox{T.~Wang\textsuperscript{63}},
\mbox{S.~Wilkins\textsuperscript{60}},
\mbox{J.~Wise\textsuperscript{64}},
\mbox{J.~Witstok\textsuperscript{23,24}},
\mbox{Z.~Wu\textsuperscript{1}},
\mbox{L.~Yung\textsuperscript{65}},
\mbox{T.~Zana\textsuperscript{2,3,22}},
\mbox{H.~Zhang\textsuperscript{66}},
\mbox{S.~Zhang\textsuperscript{38}},
M.~Zhuang\textsuperscript{54}\par\endgroup}

\newcommand{\AffiliationList}{%
\setlength{\parskip}{0pt}\setlength{\parindent}{0pt}%
\hangindent=1.4em\hangafter=1 \textsuperscript{1}Center for Astrophysics $\vert$ Harvard \& Smithsonian, 60 Garden Street, Cambridge, MA 02138, USA\par
\hangindent=1.4em\hangafter=1 \textsuperscript{2}Dipartimento di Fisica, Sapienza Universit\`a di Roma, Piazzale Aldo Moro 5, 00185 Roma, Italy\par
\hangindent=1.4em\hangafter=1 \textsuperscript{3}INAF -- Osservatorio Astronomico di Roma, Via Frascati 33, 00078 Monte Porzio Catone, Italy\par
\hangindent=1.4em\hangafter=1 \textsuperscript{4}Department of Astronomy and Astrophysics, University of Chicago, Chicago, IL 60637, USA\par
\hangindent=1.4em\hangafter=1 \textsuperscript{5}Department of Astronomy and Astrophysics, University of California, Santa Cruz, 1156 High Street, Santa Cruz, CA 95064, USA\par
\hangindent=1.4em\hangafter=1 \textsuperscript{6}Center for Computational Astrophysics, Flatiron Institute, New York, NY 10010, USA\par
\hangindent=1.4em\hangafter=1 \textsuperscript{7}Department of Astronomy, Indiana University, 727 East Third Street, Bloomington, IN 47405, USA\par
\hangindent=1.4em\hangafter=1 \textsuperscript{8}Instituto de Estudios Astrofísicos, Facultad de Ingeniería y Ciencias, Universidad Diego Portales, Av. Ejército 441, Santiago, Chile\par
\hangindent=1.4em\hangafter=1 \textsuperscript{9}Instituto de Astrof\'{\i}sica e Ci\^{e}ncias do Espa\c{c}o, Portugal\par
\hangindent=1.4em\hangafter=1 \textsuperscript{10}Faculdade de Ci\^{e}ncias da Universidade de Lisboa, Portugal\par
\hangindent=1.4em\hangafter=1 \textsuperscript{11}Department of Astronomy, School of Science, The University of Tokyo, 7-3-1, Hongo, Bunkyo, Tokyo, 113-0033, Japan\par
\hangindent=1.4em\hangafter=1 \textsuperscript{12}University of the Pacific, Stockton, CA 90340, USA\par
\hangindent=1.4em\hangafter=1 \textsuperscript{13}Univ Lyon, Univ Lyon1, Ens de Lyon, CNRS, CRAL UMR5574, F-69230, Saint-Genis-Laval, France\par
\hangindent=1.4em\hangafter=1 \textsuperscript{14}Armagh Observatory and Planetarium, College Hill, Armagh, BT61 9DG, N. Ireland, UK\par
\hangindent=1.4em\hangafter=1 \textsuperscript{15}Institute for Computational Cosmology, Department of Physics, Durham University, South Road, Durham DH1 3LE, UK\par
\hangindent=1.4em\hangafter=1 \textsuperscript{16}University of Ljubljana, Faculty of Mathematics and Physics, Jadranska ulica 19, SI-1000 Ljubljana, Slovenia\par
\hangindent=1.4em\hangafter=1 \textsuperscript{17}Department of Physics and Astronomy, University of California Davis, 1 Shields Avenue, Davis, CA 95616, USA\par
\hangindent=1.4em\hangafter=1 \textsuperscript{18}Columbia University\par
\hangindent=1.4em\hangafter=1 \textsuperscript{19}Kavli Institute for Cosmology, University of Cambridge, Madingley Road, Cambridge CB3 0HA, UK\par
\hangindent=1.4em\hangafter=1 \textsuperscript{20}Institute of Astronomy, University of Cambridge, Madingley Road, Cambridge CB3 0HA, UK\par
\hangindent=1.4em\hangafter=1 \textsuperscript{21}School of Earth and Space Exploration, Arizona State University, Tempe, AZ 85287, USA\par
\hangindent=1.4em\hangafter=1 \textsuperscript{22}INFN, Sezione di Roma I, Piazzale Aldo Moro 2, 00185 Roma, Italy\par
\hangindent=1.4em\hangafter=1 \textsuperscript{23}Cosmic Dawn Center (DAWN), Copenhagen, Denmark\par
\hangindent=1.4em\hangafter=1 \textsuperscript{24}Niels Bohr Institute, University of Copenhagen, Jagtvej 128, 2200 Copenhagen, Denmark\par
\hangindent=1.4em\hangafter=1 \textsuperscript{25}Scuola Normale Superiore, Piazza dei Cavalieri 7, I-56126 Pisa, Italy\par
\hangindent=1.4em\hangafter=1 \textsuperscript{26}Department of Astronomy, University of Michigan, 1085 S. University Ave., Ann Arbor, MI 48109, USA\par
\hangindent=1.4em\hangafter=1 \textsuperscript{27}Centre for Astrophysics Research, Department of Physics, Astronomy and Mathematics, University of Hertfordshire, Hatfield AL10 9AB, UK\par
\hangindent=1.4em\hangafter=1 \textsuperscript{28}Dipartimento di Fisica, Università di Roma Tor Vergata, Via della Ricerca Scientifica, 1, Roma 00133, Italy\par
\hangindent=1.4em\hangafter=1 \textsuperscript{29}Institute for Astronomy, University of Edinburgh, Royal Observatory, Edinburgh EH9 3HJ, UK\par
\hangindent=1.4em\hangafter=1 \textsuperscript{30}Centre for Astrophysics and Supercomputing, Swinburne University of Technology\par
\hangindent=1.4em\hangafter=1 \textsuperscript{31}Department of Physics, University of Connecticut, 2152 Hillside Road, Unit 3046, Storrs, CT 06269, USA\par
\hangindent=1.4em\hangafter=1 \textsuperscript{32}Los Alamos National Laboratory, Los Alamos, NM 87545, USA\par
\hangindent=1.4em\hangafter=1 \textsuperscript{33}NSF National Optical-Infrared Astronomy Research Laboratory, 950 N. Cherry Ave., Tucson, AZ 85719, USA\par
\hangindent=1.4em\hangafter=1 \textsuperscript{34}Institut d'astrophysique de Paris, UMR 7095, CNRS, Sorbonne Universit\'e, 98 bis boulevard Arago, 75014 Paris, France\par
\hangindent=1.4em\hangafter=1 \textsuperscript{35}IPAC, California Institute of Technology, 1200 E. California Blvd. Pasadena, CA 91125, USA\par
\hangindent=1.4em\hangafter=1 \textsuperscript{36}Institute of Physics, Laboratory for Galaxy Evolution and Spectral Modelling, EPFL, Observatoire de Sauverny, Chemin Pegasi 51, 1290 Versoix, Switzerland\par
\hangindent=1.4em\hangafter=1 \textsuperscript{37}Department of Astronomy, The University of Texas at Austin, Austin, TX 78712, USA\par
\hangindent=1.4em\hangafter=1 \textsuperscript{38}Cosmic Frontier Center, The University of Texas at Austin, Austin, TX 78712, USA\par
\hangindent=1.4em\hangafter=1 \textsuperscript{39}DTU Space, Technical University of Denmark, Elektrovej 328, 2800 Kgs. Lyngby, Denmark\par
\hangindent=1.4em\hangafter=1 \textsuperscript{40}Institute for Astronomy, University of Hawaii, 2680 Woodlawn Drive, Honolulu, HI 96822, USA\par
\hangindent=1.4em\hangafter=1 \textsuperscript{41}U.S. National Science Foundation (NSF), Alexandria, VA 22314, USA\par
\hangindent=1.4em\hangafter=1 \textsuperscript{42}Instituto de Astrof\'{i}sica de Canarias, 38205 La Laguna, Tenerife, Spain\par
\hangindent=1.4em\hangafter=1 \textsuperscript{43}ESO, Karl-Schwarzschild-Str. 2, 85748 Garching bei M\"unchen, Germany\par
\hangindent=1.4em\hangafter=1 \textsuperscript{44}Department of Physics and Astronomy, University of Maine, Orono, ME 04469, USA\par
\hangindent=1.4em\hangafter=1 \textsuperscript{45}Max Planck Institute for Astrophysics\par
\hangindent=1.4em\hangafter=1 \textsuperscript{46}Space Telescope Science Institute, 3700 San Martin Drive, Baltimore, MD 21218, USA\par
\hangindent=1.4em\hangafter=1 \textsuperscript{47}University of Toronto, Toronto, ON M5S 3H4, Canada\par
\hangindent=1.4em\hangafter=1 \textsuperscript{48}SISSA, Via Bonomea 265, 34136 Trieste, Italy\par
\hangindent=1.4em\hangafter=1 \textsuperscript{49}Physics Department, College of Science, United Arab Emirates University, PO Box 15551, Al-Ain, UAE\par
\hangindent=1.4em\hangafter=1 \textsuperscript{50}The University of Osaka\par
\hangindent=1.4em\hangafter=1 \textsuperscript{51}Department of Astronomy, Yale University, Kline Tower, 266 Whitney Avenue, New Haven, CT 06511, USA\par
\hangindent=1.4em\hangafter=1 \textsuperscript{52}Department of Physics, Yale University, P.O. Box 208121, New Haven, CT 06520, USA\par
\hangindent=1.4em\hangafter=1 \textsuperscript{53}Black Hole Initiative at Harvard University, 20 Garden Street, Cambridge, MA 02138, USA\par
\hangindent=1.4em\hangafter=1 \textsuperscript{54}Department of Astronomy, University of Illinois Urbana-Champaign, Urbana, IL 61801, USA\par
\hangindent=1.4em\hangafter=1 \textsuperscript{55}Centro de Astrobiolog\'{\i}a (CAB), CSIC-INTA, Ctra. de Ajalvir km 4, Torrej\'on de Ardoz, E-28850, Madrid, Spain\par
\hangindent=1.4em\hangafter=1 \textsuperscript{56}Konkoly Observatory\par
\hangindent=1.4em\hangafter=1 \textsuperscript{57}School of Physics and Astronomy, University of Southampton, Highfield, Southampton, SO17 1BJ, UK\par
\hangindent=1.4em\hangafter=1 \textsuperscript{58}CIERA and Department of Physics and Astronomy, Northwestern University\par
\hangindent=1.4em\hangafter=1 \textsuperscript{59}University of Arizona\par
\hangindent=1.4em\hangafter=1 \textsuperscript{60}Astronomy Centre, University of Sussex, Falmer, Brighton BN1 9QH, UK\par
\hangindent=1.4em\hangafter=1 \textsuperscript{61}Max-Planck-Institut f\"ur extraterrestrische Physik\par
\hangindent=1.4em\hangafter=1 \textsuperscript{62}INAF -- Osservatorio di Astrofisica e Scienza dello Spazio di Bologna, Via Gobetti 93/3, I-40129 Bologna, Italy\par
\hangindent=1.4em\hangafter=1 \textsuperscript{63}School of Astronomy and Space Science, Nanjing University, Nanjing, Jiangsu 210023, China\par
\hangindent=1.4em\hangafter=1 \textsuperscript{64}Georgia Institute of Technology\par
\hangindent=1.4em\hangafter=1 \textsuperscript{65}National Tsing Hua University - Taiwan\par
\hangindent=1.4em\hangafter=1 \textsuperscript{66}Department of Astronomy, School of Physics, Huazhong University of Science and Technology, Luoyu Road 1037, Wuhan, China\par
\hangindent=1.4em\hangafter=1 \textsuperscript{67}Institut f\"ur theoretische Astrophysik, Zentrum f\"ur Astronomie der Universit\"at Heidelberg, Albert-Ueberle-Str.\ 2, D-69120 Heidelberg, Germany\par
\hangindent=1.4em\hangafter=1 \textsuperscript{68}INAF -- Osservatorio Astrofisico di Arcetri, Largo E. Fermi 5, I-50125 Firenze, Italy\par
\hangindent=1.4em\hangafter=1 \textsuperscript{69}University of Oxford\par
\hangindent=1.4em\hangafter=1 \textsuperscript{70}Kavli Institute for Astronomy and Astrophysics, Peking University, Beijing 100871, China\par
}

\newcommand{\CorrespondingBlock}{\textsuperscript{\dag}Corresponding authors, who contributed equally to this work:\\
F.~Pacucci (\href{mailto:fabio.pacucci@cfa.harvard.edu}{fabio.pacucci@cfa.harvard.edu}) and R.~Schneider (\href{mailto:raffaella.schneider@uniroma1.it}{raffaella.schneider@uniroma1.it})}

\thispagestyle{genesisfirst}

{\LARGE\bfseries\color{genesisink}Black Hole Seeds and Little Red Dots: \\Connecting Formation and Growth with a JWST Multi-Cycle Treasury Program\par}
\vspace{0.8em}

{\small\AuthorList}
\vspace{0.5em}

{\small\CorrespondingBlock\par}

\vspace{1.0em}

\noindent\textbf{\color{genesisink}Abstract.}\enspace
JWST has revealed an abundant population of faint active galactic nuclei and a new class of compact, high-redshift sources known as Little Red Dots (LRDs), whose physical nature remains debated, with accreting massive black holes among the leading explanations.
The emerging demographics of early black holes, with their high inferred number densities and, in some systems, the apparent excess of black hole mass relative to stellar mass, challenge simple scenarios of black hole seeding and growth.
Despite these observational breakthroughs, the origin of early massive black holes remains uncertain, because ``black holes have no memory'': current samples cannot separate the imprint of seeding from subsequent growth. This white paper identifies the multi-cycle JWST observations needed to revolutionize our understanding of early black hole growth by: (i) distinguishing seeding pathways, (ii) constraining early black hole growth mechanisms, (iii) resolving the physical nature of broad-line emitters and LRDs, and (iv) testing black hole-galaxy co-evolution. These tasks require uniformly selected samples extending to low black hole and host-galaxy masses, allowing self-consistent measurements of black hole mass functions, Eddington-ratio distributions, and host properties. Grating spectroscopy, deep MIRI imaging, and variability analysis through repeated observations of the same fields would provide complementary tests of black hole formation and growth, their connection to host-galaxy evolution, and the nature of newly discovered populations of LRDs. In this white paper, we argue that these joint tests require a coordinated JWST Multi-Cycle Treasury program \citep{genesis}; such a program could provide the depth, uniformity, and time baseline needed to place substantially stronger, complementary constraints on early black hole formation, growth, and their connection to galaxy evolution.

\section{Scientific Justification}

JWST discoveries have challenged our understanding of how the first black holes formed, grew, and began co-evolving with their host galaxies \citep{ubler2026}. For example, the detection of broad Balmer lines with full-width at half maximum (FWHM) of $\gtrsim 1000 \, \rm km s^{-1}$ revealed accreting black holes of $\lesssim 10^6-10^8 \Msun$ at $z = 4-7$, hosted by galaxies with stellar masses of at most $10^8-10^{10} \Msun$ \citep{Kocevski2023, Harikane2023, Maiolino_2023_new, Taylor_2024}. In addition, the Little Red Dots (LRDs), compact sources with V-shaped spectral energy distributions and broad lines, outnumber UV-selected quasars of comparable luminosity by $1-2$ orders of magnitude, and they represent $> 1\%$ of all bright sources at high redshift \citep{Matthee2024, Kocevski_2024, Akins_2025}. Remarkably, in several of these newly uncovered systems, the black hole mass seems to exceed the local $\Mblack-\Mstar$ relation by factors of $\sim 10-100$ \citep{Pacucci_2023_JWST, Juodzbalis_2024_extreme, Jones_2025, Gupta_2026}. 
The current black hole frontier sits at $z \approx 10.6$ \citep{Maiolino_2023_new}, less than $500$ Myr after the Big Bang, where GN-z11 may host a $\sim 10^6 \Msun$ black hole. 

Yet current observations cannot uniquely distinguish between different seeding pathways, accretion mechanisms, and black hole–galaxy evolutionary scenarios, nor can they uniquely identify the nature of the LRDs. In particular, three debates remain wide open. 
First, black hole mass estimates may be affected by non-virial motions and radiative-transfer effects, including outflows and electron scattering in dense gas, while single-epoch virial estimators are being applied well outside their original calibration range \citep{Lupi_2024, Rusakov_2025, Pacucci_2026_DCBH, Trinca2026, Gravity2026}.
Second, we cannot yet determine whether these systems require massive seeds or can be explained by rapid growth from lighter seeds \citep{Pacucci_2022_search}. Because an overmassive black hole-to-stellar mass ratio can originate both from a heavy seed and a host whose stars formed late, its mere detection does not solve the identification problem. 
Third, the physical nature of the LRDs is uncertain.
While many LRDs display features commonly associated with accreting massive black holes, they are undetected in X-rays and radio, show limited or debated evidence for hot-dust emission, and exhibit weak or undetected variability over the available observational baselines \citep{Yue_2024_Xray, Ananna_2024,Pacucci_Narayan_2024, Inayoshi_Maiolino_2025, Madau2026a, Perez-Gonzalez_2026}.

To resolve these questions, future JWST programs should measure, for the same sources: (i) black hole masses, (ii) accretion diagnostics, and (iii) host-galaxy properties. Several elements, including their redshift distribution, abundances, clustering, and host-galaxy environments, all evaluated through a consistent selection function, would constrain seeding, accretion, and stellar assembly, enabling comparisons with theoretical models.
These multi-cycle observations would also establish the nature of LRDs: their abundance, measured consistently across redshift ranges, and their variability could trace either existing black holes or an earlier stage of seed formation. We argue that a JWST multi-cycle program should combine deep imaging, spectroscopy, and repeated observations within a common parent population across the first billion years, with well-characterized selection and completeness.

Below we outline four coupled science cases where these constraints will have the greatest impact.

\section{Key Science Cases}

\subsection{The origin of massive black hole seeds}
\label{sec:bh_seeding}

The origin of the first massive black holes (BHs) remains a major open question in early structure formation. Proposed pathways range from light seeds ($\sim 10^2\,M_\odot$),
originating from Population III stellar remnants, to intermediate- and heavy-mass seeds ($\sim 10^3$--$10^5\,M_\odot$), produced through stellar dynamical processes, runaway collisions, or the collapse of supermassive stars formed in rapidly accreting gas clouds \citep{Inayoshi2020,Volonteri2021,Regan2024}. An additional, intriguing possibility is that the seeds are primordial black holes (PBHs). This option was revived to explain the JWST overmassive population but remains tightly constrained by the cosmic microwave background for masses above $\sim 10^4 \Msun$ \citep{Dayal2024, Zhang_2025_PBH, Prole2025}.
The relative importance of these channels is uncertain and depends on the primordial gas and its thermal, chemical, and dynamical evolution.

Recent simulations have dramatically expanded the range of environments in which we believe that massive seeds may form. Internal radiative feedback can trigger direct collapse in atomic-cooling haloes \citep{Chiaki2023,Kiyuna2026}, while supermassive stars and dense stellar clusters may form even in slightly metal-enriched environments \citep{Chon2020,Chon2025}. Additionally, cosmological simulations suggest that rapid galaxy assembly may favor massive seed formation \citep{Prole2026c}.

These pathways lead to different seed properties, especially regarding their initial mass function, occupation fractions, and host-galaxy environments. However, subsequent accretion and mergers can partially erase the initial signatures of the seed population. Semi-analytical models and cosmological simulations, including CAT, DELPHI, BRAHMA, SEEDZ, and AESOPICA, have explored how different seeding prescriptions translate into the evolving BH and host-galaxy properties \citep{Valiante2016,Trinca2022,Trinca2023,Dayal2019, Bhowmick2024a,Bhowmick2026a, Koudmani_2026,Prole2026a,Mehta2026}.

\textbf{What could a JWST MCT program do?}
Distinguishing between seeding scenarios requires going beyond the rare, luminous AGN accessible today, with different observables in two redshift regimes. At $z \approx 5-7$, the imprint of the seed mass function survives below $M_{\rm BH} \sim 10^6 \Msun$, where heavy-seed models predict a deficit of BHs and light-seed models a continuous distribution \citep{Trinca2022, Bhowmick2024a, Koudmani_2026}: the bolometric luminosity function must reach $L_{\rm bol} \approx 3 \times 10^{43}$ erg s$^{-1}$ (a $10^6 \Msun$ black hole at a quarter of the Eddington rate), together with the active fraction in hosts of $M_\star \approx 10^8 \Msun$. At $z \gtrsim 9$ only the bright end is accessible: the number density of $10^6$--$10^7 \Msun$ BHs near the Eddington rate ($L_{\rm bol} \gtrsim 10^{44}$ erg s$^{-1}$) is sufficient to probe the properties of the seeds, because light seeds growing continuously at the Eddington rate cannot have produced them in the time available, although earlier super-Eddington episodes could relax this constraint. Recent JWST observations have begun to uncover such low-mass BHs \citep{Geris2026}. Our JWST MCT program could turn these detections into a complete census (first regime) and a volume-limited count (second regime).

\subsection{Early black hole growth and accretion physics}

Constraining the properties of early black hole seeds is, fundamentally, a question of time constraints.
A black hole accreting at the Eddington rate (assuming radiative efficiency $\epsilon = 0.1$) e-folds its mass every $\sim 50$ Myr, so the budget between a seed formed at $z = 25$ \citep{BL01} and the quasar with a BH mass of $1.6 \times 10^9\,{\rm M_\odot}$  at $z = 7.64$ \citep{Wang_2021_quasar} is $\approx 11$ e-folding times, which is sufficient for a $10^5\,{\rm M_\odot}$ seed. However, the same (Eddington-limited) growth period is insufficient for a $100 \,{\rm M_\odot}$ seed, which would stop below $10^7\,{\rm M_\odot}$. GN-z11 poses the same problem at $z = 10.6$, where the budget shrinks to $\approx 6$ e-folds: a light seed could not have reached $10^5\,{\rm M_\odot}$, while the black hole inferred from its spectrum has a mass of $\sim 10^6\,{\rm M_\odot}$ and is estimated to be accreting at several times the Eddington rate \citep{Maiolino_2023}. 

One possibility to overcome these issues is super-Eddington growth \citep{Pezzulli_2016, Inayoshi2020, Volonteri2025}. In slim-disk scenarios, photon trapping efficiently lowers the radiative efficiency, allowing the BH mass to grow faster, so a few episodes at $ \lambda_{\rm Edd} \gtrsim 1$ may be enough to close the gap from stellar-mass seeds \citep{Madau_2014_super, Volonteri_2015, Pacucci_2015, Lupi2016}. However, sustaining rapid gas inflow over long timescales is challenging. In fact, feedback starves stellar-mass seeds in low-mass halos \citep{Alvarez_2009, Milosavljevic_2009}, and even heavy seeds grow efficiently only within a narrow range of seed and host properties \citep{Pacucci_2017}.

Cosmological simulations and semi-analytical models show that
the efficiency and duty cycles of early BH accretion depend
strongly on the interplay between gas supply, stellar and AGN
feedback, and their evolving environment
\citep{Lupi2016,Pezzulli_2017,Schneider2023,Lupi_2024}.
Consequently, the observed abundance of massive BHs at high $z$ cannot be uniquely mapped onto their initial seed
masses without properly constraining the accretion histories.

Growth at $\lambda_{\rm Edd} \lesssim 1$ shifts the black hole mass function smoothly in mass with time, whereas episodic super-Eddington growth produces, at any epoch, a majority of dormant BHs and a minority caught in a burst \citep{Pezzulli_2017, Trinca2022, Schneider2023, Trinca2026, Zana_2026}. For example, the $4 \times 10^8\,{\rm M_\odot}$ black hole at $z = 6.68$ accreting at $2\%$ of the Eddington rate \citep{Ignas_2024_overmassive} is the first clear example of what several studies define as the dormant majority of high-redshift BHs. The Eddington-ratio distribution, and its link to electromagnetic properties, is also an important prediction. Super-Eddington flows are expected to be X-ray weak \citep{Pacucci_Narayan_2024, Lambrides_2024, Maiolino_2024_Xray} and, if the inflow is chaotic, to keep the spin low \citep{King_2006}, thus lowering $\epsilon$ and accelerating growth further. 
Theoretical models also differ in their predictions for the faint-end AGN luminosity function and on how the active fraction depends on host mass \citep{Koudmani_2026, Caleno_2026}.

\textbf{What could a JWST MCT program do?} 
The distributions of several physical parameters predicted by these theoretical models separate these prescriptions. Of course, distributions require a mass-complete sample rather than a set of candidates, observed at very specific masses. 
A JWST MCT program would be essential to extend such samples to the lowest masses and luminosities. 
In this case, the black hole mass function and the Eddington-ratio distribution at $z = 4$--$9$ would constrain the duty cycle, provided that the program consistently reaches black hole masses of $M_\bullet \sim 10^6\,{\rm M_\odot}$.
The fraction of X-ray-weak sources, as well as the incidence and kinematics of outflows, would provide complementary tests of super-Eddington accretion and feedback as a function of $\lambda_{\rm Edd}$. The same measurements at $z \gtrsim 9$ would provide important timing constraints on seeding scenarios, especially when connected to the Eddington-ratio distribution described above.
The detection of massive BHs accreting at low Eddington
ratios at very early epochs would place stringent
constraints on their previous growth histories \citep{Pacucci_2022_search}, but would not by itself establish the need for heavy seeds,
since earlier episodes of rapid or super-Eddington
accretion may have played a decisive role
\citep{Madau_2014_super,Lupi2016,Schneider2023,Trinca2026}.

\subsection{The physical nature of Little Red Dots}

Broad Balmer lines (FWHM $\sim 1000$--$4000$\,km\,s$^{-1}$) in many spectroscopically confirmed LRDs suggest accreting BHs \citep{Matthee2024, Greene_2023}, yet their X-ray weakness and debated hot-dust signatures challenge conventional AGN interpretations \citep{Yue_2024_Xray, Ananna_2024, Maiolino_2024_Xray, Casey_2024}.

Four main models are debated: (i) accreting BHs, possibly super-Eddington, embedded in gas whose geometry and optical depth shape the observed emission, potentially unifying LRDs and Little Blue Dots \citep{Pacucci_Narayan_2024, Madau2026a, Madau2026LBD}; (ii) BHs embedded in dense, reprocessing envelopes, including ``black hole stars'' and late-stage quasi-stars \citep{Naidu_2025a, deGraaff_2025_Cliff, Begelman_2008, Begelman_2026}; (iii) very massive or supermassive stars before black hole formation \citep{Hosokawa_2013, Nandal_2026}; and (iv) dense stellar clusters in formation \citep{Chisholm_2026}. These model categories also differ in the BH phase they describe. Models (i) and (ii) contain BHs already in place that may retain signatures of their seeds, whereas models (iii) and (iv) may probe conditions preceding black hole formation. Simulations predict short LRD-like phases for massive seeds \citep{Chon_2026_Nature}, while other models associate LRDs with overmassive BHs embedded in dense gas \citep{LaChance_2025}.

Four complementary diagnostics can help distinguish these scenarios. Broad-line profiles probe gas kinematics and radiative-transfer effects (e.g., electron-scattering wings) which complicate virial BH mass estimates \citep{Rusakov_2025,Madau2026c}. Rest-frame near-infrared emission constrains hot dust, stellar light, and cool photospheres ($T_{\rm eff} \approx 4000-6000$ K), although the latter cannot distinguish an embedded black hole from a supermassive star \citep{Perez-Gonzalez_2026, Hosokawa_2013, Chen_2026_spectral, Begelman_2008}. Variability probes the size and structure of the emitting region: extended photospheres may suppress rapid fluctuations, whereas stellar clusters can vary through individual massive stars or transients. To conclude, spectropolarimetry provides an additional diagnostic of scattering geometries and obscuration \citep{Madau2026d}.

Remarkably, LRD demographics provide independent constraints. LRDs occur in roughly $1\%$ of galaxies and are most common at $z \approx 4.5-8$ \citep{Kocevski_2024, Pacucci_Loeb_2025}. Their abundance is the product of two unknowns: the duration of the LRD phase and the fraction of galaxies that go through it. With LRDs at $\sim 1\%$ of galaxies over the $\sim 700$ Myr between $z = 8$ and $z = 4.5$, a supermassive-star phase of $\sim 3$ Myr \citep{Woods_2019} would have to recur $\approx 2-3$ times in every galaxy; a $\sim 20$ Myr quasi-star phase \citep{Begelman_2008} would need one galaxy in three; and $\sim 230$ Myr of Eddington-limited growth, only a few percent of them. The apparent cutoff of the LRD luminosity function near $L_{\rm bol} \sim 10^{45}$ erg s$^{-1}$ further constrains black hole or envelope masses \citep{Ma_2026_cutoff}.

\textbf{What could a JWST MCT program do?}
Existing tests are limited by small spectroscopic samples, heterogeneous MIRI depths, and short variability baselines \citep{Rusakov_2025}. A multi-cycle program, combining deep MIRI imaging with medium-resolution spectroscopy, would uniformly apply these diagnostics. These measurements, coupled with redshift-dependent abundances and completeness, could constrain the relative contributions of different LRD populations and establish whether some trace the formation or early growth of massive BHs.

\subsection{Black hole--galaxy co-evolution}

In the local Universe, BHs weigh $\approx 0.1\%$ of their host's stellar mass \citep{Kormendy_Ho_2013, Reines_Volonteri_2015}. 
At $z = 4$--$7$, some JWST-detected broad-line AGN appear to lie $10$--$100$ times above that relation, with offsets exceeding $3\sigma$ in some analyses \citep{Pacucci_2023_JWST}. Remarkably, the offset grows with redshift \citep{Gupta_2026} and, in some sources, the black hole may outweigh the stars \citep{Furtak_2023, Juodzbalis2026}.

Several issues could compromise these measurements. First, flux-limited samples select BHs on the high side of the intrinsic scatter \citep{Lauer_2007}, so the local relation could reproduce an offset of this size if the scatter is larger than expected \citep{Somerville_2009, Li_2024_iceberg,Koudmani_2026}. The stellar masses of hosts dominated by a point source are highly uncertain, and for most LRDs the host is undetected \citep{Chen_2025, Durodola_2024}. 
In addition, single-epoch virial BH mass estimates carry substantial systematic uncertainties. On the theory side \citep{Volonteri2025}, the offset has several possible interpretations: (i) a signature of heavy seeds, which start their life overmassive and lose that memory on a timescale set by the growth of the host \citep{Agarwal_2013, Natarajan_2017, Visbal_Haiman_2018, Pacucci_2024_z_evolution, Inayoshi_2024, Scoggins_2023, Mehta2026}. In this scenario, the largest offsets occur near the seeding era, at the lowest masses and highest redshifts; (ii) a transient departure from a relation regulated by feedback, to which hosts converge as they grow \citep{Silk_1998, King_2003, Habouzit_2017}; or (iii) scatter and a selection bias above a relation close to the local one, which is where most large-volume simulations place their BHs at $z \sim 6$ \citep{Weller_2023, Koudmani_2026, Davari_2026}.

\textbf{What could a JWST MCT program do?} A JWST MCT program could provide host masses from imaging and continuum spectroscopy that are deep enough for the hosts of LRDs to be directly detected, and dynamical masses from integral-field spectroscopy for a subset of such BHs \citep{Ubler_2023, Maiolino_2023_new}, which provide crucial information regarding the virial calibration and the $M_{\rm BH}$--$\sigma$ test. The models \citep[e.g.][]{Somerville2008, Trinca2022, Koudmani_2026} can then be processed through the same selection function and compared with the observations on a like-for-like basis.

\subsection{Executive summary}

The four science cases detailed above converge on one requirement: obtaining black hole masses, accretion diagnostics, and host-galaxy properties for the same sources, drawn from a uniformly selected parent population and revisited across multiple cycles. No combination of independently conceived and executed GO programs can provide these data. With such a uniform sample, the community could make significant advances in understanding the nature of the LRDs, while the Eddington-ratio distribution would show how the first massive BHs grew, and the $\Mblack$--$\Mstar$ relation would rest on a well-characterized population. The high-$z$ community would then have the data to measure black hole growth over the first billion years in a complete, systematic way.




\section{Affiliations}
{\footnotesize\itshape\textcolor{genesisgray}{\AffiliationList}}

\bibliographystyle{unsrtnat-short} 
\bibliography{rss, FP}

\end{document}